\documentclass[aps,prd,nofootinbib,twocolumn,showpacs]{revtex4-1}
\usepackage {graphicx}
\usepackage {hyperref}
\usepackage {color}
\usepackage{textcase}
\usepackage{amsmath, amsthm, amsfonts}
\usepackage{enumerate}
\usepackage{soul}
\usepackage[normalem]{ulem}
\usepackage[english]{babel}
\usepackage[T1]{fontenc}
\usepackage{graphicx}
\usepackage{amsmath}
\usepackage{hyperref}
\usepackage{midfloat}
\usepackage{lineno}

\newcommand{\be}{\begin{equation}}
\newcommand{\ee}{\end{equation}}

\begin{document}
\title{Revised predictions of neutrino fluxes from Pulsar Wind Nebulae}

\author{Irene Di Palma}
\email{Irene.DiPalma@roma1.infn.it}
\affiliation{Istituto Nazionale di Fisica Nucleare, Sezione di Roma, Italy}
\affiliation{Universit\'{a} di Roma La Sapienza, I-00185 Roma, Italy}

\author{Dafne Guetta}
\email{dafne.guetta@oa-roma.inaf.it}
\affiliation{Osservatorio astronomico di Roma, v. Frascati 33, 00040 Monte Porzio Catone, Italy}
\affiliation{Department of Physics
	Optical Engineering, ORT Braude, P.O. Box 78, Carmiel, Israel}
%\affiliation{Department of Physics, Technion, Israel}

\author{Elena Amato}
\email{amato@arcetri.astro.it}
\affiliation{Istituto Nazionale di Astrofisica}
\affiliation{Osservatorio astrofisico di Arcetri,  Firenze, Italy}

%\date{} % Activate to display a given date or no date (if empty),
         % otherwise the current date is printed 
%\linenumbers

\begin{abstract}
Several Pulsar Wind Nebulae (PWNe) have been detected in the TeV band in the last decade.The TeV emission is typically interpreted in a purely leptonic scenario, but this usually requires that the magnetic field in the Nebula be much lower than the equipartition value and the assumption of an enhanced target radiation at IR frequencies. 
In this work we consider the possibility that, in addition to the relativistic electrons, also relativistic hadrons are present in these nebulae. Assuming that part of the emitted TeV photons are of hadronic origin, we compute the associated flux of $\sim 1-100$ TeV neutrinos. We use the IceCube non detection to put constraints on the fraction of TeV photons that might be contributed by hadrons and estimate the number of neutrino events that can be expected from these sources in ANTARES and in KM3Net.
\end{abstract}

\maketitle

\section{Introduction}

Pulsar Wind Nebulae (PWN) are diffuse nebulae of non-thermal radiation associated with the presence of a highly spinning, strongly magnetised neutron star. The central star of a PWN might be detected as a pulsar or not, in both cases it must emit what is called a pulsar wind, namely a relativistic magnetised outflow, mainly made of electron-positron pairs. Confinement of this outflow by the surrounding Supernova Remnant or by the interstellar medium (ISM) leads to the formation of a termination shock, at which the wind is slowed down to non-relativistic bulk speed and particles are accelerated to a power-law (or broken power-law) distribution. The interaction of these highly energetic leptons with the ambient magnetic field and with the radio and IR background radiation is thought to be at the origin of the nebular emission, from radio wavelengths to the TeV band. 

The material component of the wind is mainly made of electrons and positrons due to the pair production process that takes place in the star magnetosphere, but  the presence of a minority of ions cannot be excluded. In addition, in spite of being a minority, ions could still carry a dominant fraction of the total wind energy, as we will  explain in the following.

The theoretical upper limit on a possible hadronic component is given by the Glodreich and Julian particle flux
\be
\dot N_{\rm GJ}=\frac{8 \pi^2 B_* R_*^3}{Z e c P_*^2}=5 \times 10^{34}\ s^{-1}\ B_{12}\ R_{10}^3\ P_{-2}^{-2}\ Z^{-1}
\label{eq:ngj}
\ee
where $B_{12}$ is the magnetic field at the star surface ($B_*$) in units of $10^{12}$G, $R_{10}$ is the star radius ($R_*$)  in units of 10 km, $P_{-2}$ is the star rotation period ($P_*$) in units of $10^{-2}$ s, $c$ is the speed of light and $Ze$ the electric charge of the nucleus. The electron-positron pairs in the wind will be $\kappa$ times with $\kappa$ the pair multiplicity of the neutron star, namely the number of pairs produced in the magnetosphere by each electron directly emitted from the star surface. 

Before the termination shock, the pulsar wind can be described as a cold relativistic outflow, carrying the entire spin down power of the star, $\dot E$:
\be
L_{\rm wind}=\dot E_{\rm NS}=\dot N_{\rm GJ} \left(m_i +\kappa m_e\right) \Gamma_w c^2+L_B
\label{eq:winden}
\ee
where $\Gamma_w$ is the wind bulk Lorentz factor and $L_B$ is the wind magnetic luminosity, which is usually assumed to be less than that associated with particles, $m_i$ and $m_e$ are the ion and electron mass respectively. Focusing on the matter content of the wind in terms of energy it is immediately clear that ions can carry most of the wind energy even if they are fewer than leptons by number. This is what will happen if the pulsar multiplicity is $\kappa < m_i/m_e\approx 2000 A$, with $A$ the mass number. Current theoretical estimates of $\kappa$ are in the range $10^2-10^4$, so that an hadronic component, if present, could indeed dominate the energy content of the wind.

The possible presence of hadrons could also help solving one of the most mysterious aspects of the physics of PWNe, the question of what is the mechanism responsible for particle acceleration in these sources. As already mentioned, the acceleration process is thought to take place at the pulsar wind termination shock, which however is a highly relativistic magnetised shock, the most hostile environment in Nature for efficient particle acceleration. Yet in several PWNe particles are inferred to be accelerated to several hundreds TeV (up to 1 PeV, in the Crab Nebula, close to the "knee" in the Cosmic Ray spectrum) and with very high efficiency, up to 30\%. One of the very few mechanisms that have been proposed to explain such efficient acceleration requires that most of the wind energy be carried by hadrons: at the termination shock these would radiate a large fraction of their energy through the emission of waves that can be efficiently absorbed by the pairs and lead to their acceleration.

Proofs of the presence of relativistic hadrons in PWNe have been so far elusive, as for most astrophysical sources. In this particular environment the most likely way in which hadrons would show their presence is through nuclear interactions that would lead to pion production. The following decay of neutral and charged pions would produce $\gamma$-rays and neutrinos, respectively,  in the TeV range.

PWNe are the most numerous sources of TeV photons in the Galaxy. Tens of such objects have been detected by TeV telescopes in the last decade, and many of the yet unidentified sources of GeV and TeV photons are suspected to be associated with a PWNe. 

The detected TeV emission is usually interpreted as the result of Inverse Compton Scattering (ICS) of the external radiation by the highly energetic electrons and positrons whose synchrotron emission is at the origin of the nebular luminosity at lower frequencies, from radio to X-rays and sometimes up to 10-100 MeV $\gamma$-rays. Reproducing the TeV emission within a purely leptonic model, however, often requires uncomfortable assumptions: the TeV flux often happens to be so large that in order to interpret it as ICS one needs to invoke a very weak magnetic field in the nebula, much below equipartition, and/or enhanced radio-IR background. These requirements would be largely alleviated if part of the TeV $\gamma$-rays were of hadronic origin.

In this work, we consider the possibility that part of the TeV emission is due to the decay of neutral pions produced in nuclear collisions. The same process that produces the neutral pions and subsequently the TeV photons would also generate charged pions that decay into neutrinos of similar energy.
The IceCube high-energy neutrino telescope has been collecting data since 2006 and so far no neutrino event has been associated with a PWN. However, some of the TeV detected nebulae, according to the estimates of expected neutrino fluxes based on the observed photon fluxes, are very promising already for the ANTARES detector and further more for the upcoming neutrino Telescope KM3Net. We use the IceCube non detection to put constraints on the fraction of TeV photons that might be contributed by hadrons and estimate the number of neutrino events that can be expected from these sources in ANTARES and in KM3Net.

\section{Neutrino Telescopes}
\begin{figure}[h]
\begin{center}
\includegraphics[width=0.5\textwidth]{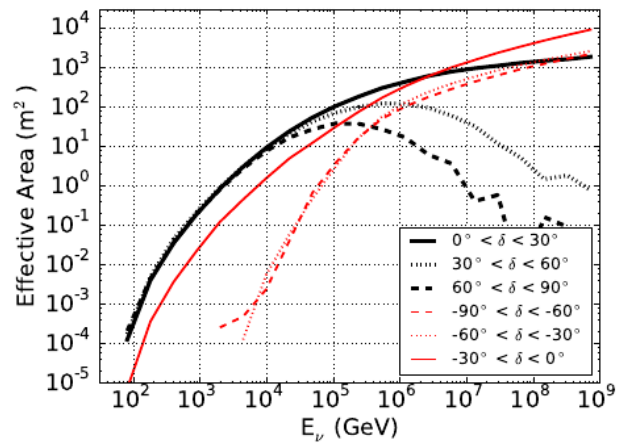}
\caption{Neutrino effective area for the 86-string detector as a function of primary neutrino energy for six declination bands. 
The effective area is the average of the area for $\nu_{\mu}$ and $\bar{\nu}_\mu$, \cite{IcePoint}.}
\label{Iceeffarea}
\end{center}
\end{figure}

The high-energy neutrinos interact with nucleons present in the detector producing secondary particles, which travel faster than the speed of light in the sea or ice and threrefore induce the emission of Cherenkov light.
These photons are detected by optical sensors deployed in sea or ice. In the following we briefly describe the basic characteristics of each telescope considered in this work.\\
IceCube is a neutrino detector located at the geographic South Pole \cite{detIce}.  In the final detector configuration, the digital optical modules, deployed in the Antartic ice, are arranged on 86 vertical strings of 60 sensors each, spread over depths between 1450m and 2450m with vertical distances of 17 m between sensors. Seventy-eight strings have a horizontal spacing of about 125 m and cover a hexagon with a surface area of roughly 1 $km^2$. Eight additional strings together with the seven surrounding IceCube strings form the more densely instrumented central DeepCore detector \cite{Ice2, Ice3}. The IceCube detector was completed in December 2010. Figure \ref{Iceeffarea} displays the effective area of the 86 strings configuration of the IceCube detector as a function of the neutrino energy for different values of declination.\\
%\subsection{IceCube}
%IceCube:
%The IceCube effective area curves for negative declinations (overhead at the South Pole) continue to rise with neutrino energy. 
%This is demonstrated in Figure~\ref{weighted_Aeff}, where the IceCube effective area is averaged over declination and the expected number of neutrinos per logarithmic energy bin is plotted, assuming an $E_\nu ^{-2}$ spectrum.

\begin{figure}[h]
\begin{center}
\includegraphics[width=0.5\textwidth]{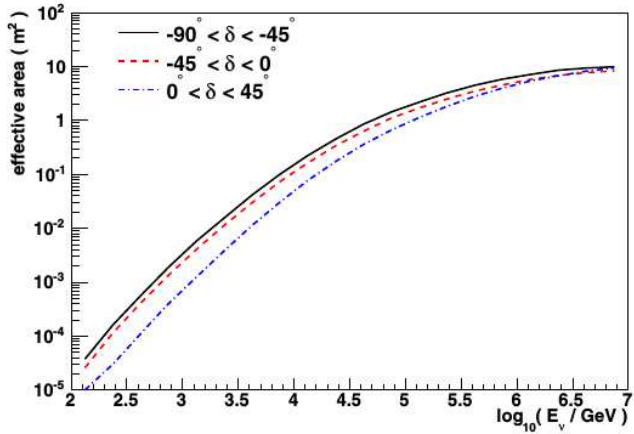}
\caption{The figure displays the effective area of the ANTARES detector for point sources as a function of the neutrino energy for different declinations, \cite{ANT}. }
\label{ANTeffarea}
\end{center}
\end{figure}

The ANTARES detector is currently the only deep sea high energy neutrino telescope, which is operating in the Northern hemisphere \cite{detAN}. The telescope covers an area of about 0.1 $km^2$ on the sea bed, at a depth of 2475 m, 40km off the coast of Toulon, France. In its full configuration, it is composed of 12 detection lines, each comprising up to 25 triplets of photomultiplier tubes \cite{AN2}, each triplet is located in one of the storeys, regularly distributed along 350 m, the first storey being located 100 m above the sea bed.The telescope reached its nominal configuration, with 12 lines immersed and taking data, in May 2008.
Figure \ref{ANTeffarea} shows the effective area of the ANTARES detector, with selection and reconstruction criteria optimized for the search of  point like sources, as a function of the neutrino energy for different declinations, \cite{ANT}.
ANTARES is planned to be followed by a multi-cubic-kilometer detector in the Mediterranean sea called KM3NeT in the next few years.
%Two of these high-energy neutrino observatories are currently in operation: IceCube \ref{detI-Ice3} – a cubic-kilometer detector at the geographic South Pole – and Antares [20] in the Mediterranean sea. Antares is planned to be followed by a multi-cubic-kilometer de- tector in the Mediterranean sea called KM3NeT in the following years [21]. 

\begin{figure}[h]
\begin{center}
\includegraphics[width=0.56\textwidth]{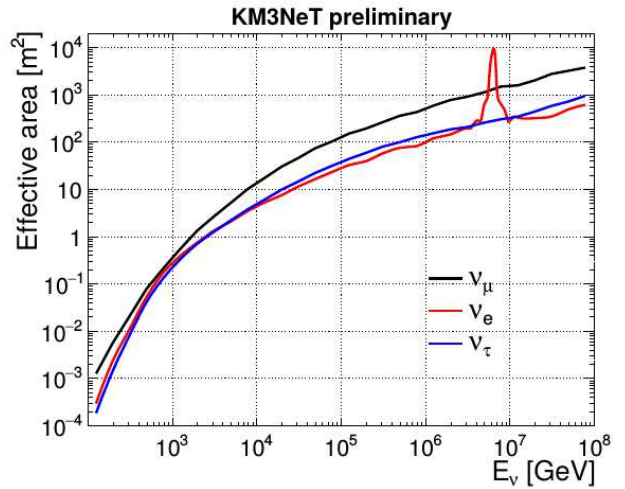}
\caption{This plot is taken from \cite{detKM3}, it shows the effective areas of ARCA (two blocks) at trigger level for $\nu_{\mu}$, $\nu_{e}$ and $\nu_{\tau}$, as a function of neutrino energy $E_{\nu}$. The effective area is defined relative to an isotropic neutrino flux incident on the Earth, is
averaged over both $\nu$ and $\bar{\nu}$, and includes both NC and CC interactions. The peak at 6.3 PeV is due to the
Glashow resonance of $\bar{\nu}_{e}$.}
\label{km3effarea}
\end{center}
\end{figure}

KM3Net \cite{detKM3} is the future generation of under water neutrino telescope. 
The infrastructure will consist of three so-called building blocks, each made of 115 strings of 18 optical modules, that have 31 photo-multiplier tubes each. KM3Net is made of KM3Net/ARCA and KM3Net/ORCA. The former will consists of two building blocks that will be deployed at 3500m depth in a site 80 km SE  of Porto Palo di Capo Passero, Sicily, Italy, while the third one, called KM3Net/ARCA, will be located at 2200m depths in a site close to ANTARES (Toulon), France. KM3Net/ARCA has large spacing between strings to target astrophysical neutrinos at TeV energies, whereas the aim of KM3Net/ORCA is to determine fundamental properties of neutrinos. Fig. \ref{km3effarea} shows the effective area of KM3Net/ARCA as a function of the different neutrino flavours, $\nu_{\mu}$, $\nu_{e}$ and $\nu_{\tau}$, at the trigger level \cite{detKM3}. The calculation of the effective area includes the detector geometric acceptance but not the efficiency of the apparatus and the effects of selection and reconstruction criteria.

\section{TeV observations of Pulsar Wind Nebulae}
During the last few years, the number of PWNe detected at TeV energies has increased from 1 \cite{trevor} to 28. The  current number of detected nebulae, mostly contributed by the H.E.S.S. survey of the Galactic plane \cite{car}, is similar to the number of these sources that are characterised at other frequencies.
The Cherenkov Telescope Array \cite{actis} will likely increase this number to several hundreds, probably providing an essentially complete account of TeV emitting PWNe in the Galaxy. 

In Table~\ref{tb:PWNlist} we list a collection of PWNe that have been detected at multi-TeV photon energies, indicating for each of them the source declination, $\delta$, the photon flux at 1 TeV, $N_0$, and the spectral index at TeV energies, $\alpha_{\gamma}$, according to a description of the TeV photon spectrum in the form:
\be
\frac{dN_\gamma}{dE_\gamma}=N_0\left(\frac{E_\gamma}{\rm 1 TeV}\right)^{-\alpha_\gamma}\ ,
\label{eq:TeVspec}
\ee
where $dN_\gamma/dE_\gamma$ is the number of photons per unit energy interval, time, surface and steradian.

As we mentioned in the introduction, a possible mechanism to interpret the TeV fluxes is the Inverse Compton Scattering (ICS), on the ambient photon field, of the electrons responsible for the synchrotron emission at lower frequencies. The target radiation for ICS might be composed in general by the sum of two different contributions: the external photon field and the internal radiation produced by the source itself, typically lower energy synchrotron emission by the same population of accelerated particles. In reality, however, the only source for which the internally produced photon field plays an important role is the Crab Nebula, which is absolutely exceptional for its brightness as a synchrotron emitter. In all other cases, the photons that are upscattered in energy belong to the CMB or to the IR background. The latter is typically taken as the sum of two diluted black body spectra peaking in the Far Infrared, at $T_{\rm FIR}\approx 25$ K and in the Near Infrared, at $T_{\rm NIR}\approx 3000$ K \cite{porter06}. The energy density in the two components varies depending on location in the Galaxy. 

Let us now focus on the photons belonging to the FIR field, since this is likely to provide the dominant contribution for most sources (see e.g. \cite{torres}). The typical Lorentz factor that electrons must have in order to upscatter these photons to TeV energies is 
\be
\gamma_e \approx \sqrt{\frac{1 TeV}{k_B T_{\rm rad}}}\approx 2 \times 10^7 \left(\frac{T_{\rm rad}}{25{\rm K}}\right)^{-1/2}\ .
\label{eq:nuics}
\ee
Their synchrotron emission will be then observed in the X-rays for a typical strength of the nebular magnetic field in the 100 $\mu$G range:
\be
h \nu_{\rm sync}\approx \frac{h}{6\pi} \frac{e B}{m_e c} \gamma_e^2 \approx 0.2\ {\rm keV} \left(\frac{B}{100 \mu{\rm G}}\right)\ \left(\frac{T_{\rm rad}}{25{\rm K}}\right)^{-1}\ .
\ ,
\label{eq:nusync}
\ee

The ratio between the power emitted as ICS radiation and that emitted as synchrotron radiation by the same electrons provides an estimate of the ambient magnetic field if the energy density of the target radiation is known:
\be
B_{\rm ICS}\approx 3 \mu {\rm G} \left(\frac{w_{\rm rad}}{\rm eV}\right)^{1/2} \sqrt{\frac{L_{\rm sync}}{L_{\rm ICS}}}\ .
\label{eq:Bics}
\ee

An independent estimate of the magnetic field can be obtained from the measurement of synchrotron emission alone if some assumption is made for the ratio between magnetic and particle energy density in the nebula, $w_{\rm B}$ and $w_{\rm part}$ respectively, hereafter. In general the synchrotron emissivity of a source depends on the combination of $w_{\rm B}$ and $w_{\rm part}$. An assumption that is commonly made to disentangle the two is that of equipartition or minimum energy: one assumes that the source host particles and field with equal sharing of its total energy between the two, a condition that approximately also guarantees that the total energy in the source is the minimum compatible with the observed synchrotron luminosity.

The determination of $B$ is as follows. 
Let us assume that the source hosts a non-thermal particles distribution in the form
\be
N(\gamma)=K_e \gamma^{-p}
\label{eq:ngamma}
\ee
where $N(\gamma)$ is the number of particles per unit energy interval integrated over the entire nebular volume.
The total energy in particles will be:
\be
W_{\rm part}=m c^2 \int_{\gamma_m}^{\gamma_M} \gamma N(\gamma) d\gamma=m c^2 \frac{K_e}{2-p}\left[\gamma_M^{2-p}-\gamma_m^{2-p}\right]
\label{eq:Ep}
\ee

The synchrotron emissivity (energy per unit time per unit frequency interval) associated to this particle distribution can be written as:
\be
S_\nu=\frac{a_1}{2 a_2} K_e\ B \left(\frac{\nu}{a_2 B}\right)^{-\frac {p-1}{2}}
\label{eq:sync}
\ee
where, from synchrotron theory, assuming emission only at the characteristic frequency $\nu_c=a_2 B \gamma^2$,
\begin{eqnarray}
a_1=\frac{1}{9 \pi} c \sigma_T \ \ \ & \ \ \   {\rm and}   \ \ \ &  \ \ \ a_2=0.29 \frac{3}{2} \sqrt{\frac{2}{3}} \frac{e}{mc}\\ .
\label{eq:a1a2}
\end{eqnarray}

The total emissivity, integrated over a frequency range [$\nu_m$, $\nu_M$] then reads:
\be
S_{\rm TOT}=\frac{a_1}{2a_2}\frac{K_e}{3-p} B^2\left[\left(\frac{\nu_M}{a_2 B}\right)^{\frac{3-p}{2}}-\left(\frac{\nu_m}{a_2 B}\right)^{\frac{3-p}{2}}\right]
\label{eq:stot}
\ee

From the latter equation we can express $K_e$ in terms of the emission parameters and magnetic field as:
\be
K_e=\frac{3-p}{a_1} \frac{S_{\rm TOT} }{B^2} \left[\left(\frac{\nu_M}{a_2 B}\right)^{\frac{3-p}{2}}-\left(\frac{\nu_m}{a_2 B}\right)^{\frac{3-p}{2}}\right]^{-1}\ .
\label{eq:kp}
\ee
By use of this expression we can then rewrite the nebular energy content in particles (Eq.~\ref{eq:Ep}) as:
\be
W_{\rm part}=\frac{3-p}{2-p}\ \frac{mc^2}{a_1}\ \frac{S_{\rm TOT}}{B^2} \
\frac{\left(\frac{\nu_M}{a_2 B}\right)^{\frac{2-p}{2}}-\left(\frac{\nu_m}{a_2 B}\right)^{\frac{2-p}{2}}} 
{\left(\frac{\nu_M}{a_2 B}\right)^{\frac{3-p}{2}}-\left(\frac{\nu_m}{a_2 B}\right)^{\frac{3-p}{2}}} .
\label{eq:Epkp}
\ee
where we have used in Eq.~\ref{eq:Ep} the expression of the characteristic synchrotron frequency to replace $\gamma_M$ and $\gamma_m$.

Finally we rewrite Eq.~\ref{eq:Epkp} in a way that highlights the dependence on $B$:
\be
W_{\rm part}=\frac{3-p}{2-p}\ \frac{a_2^{1/2}}{a_1} \frac{mc^2}{B^{3/2}} \frac{\nu_M^{\frac{2-p}{2}}-\nu_M^{\frac{2-p}{2}}}
{\nu_M^{\frac{3-p}{2}}-\nu_M^{\frac{3-p}{2}}}\ S_{\rm TOT}
\label{eq:Epfin}
\ee
At this point we can write the total energy in particles and fields corresponding to the emission $S_{\rm TOT}$ as:
\be
W_{\rm TOT}=V_N \frac{B^2}{8 \pi} \left(1+\xi_B B^{-7/2}\right)
\label{eq:wtot}
\ee
with $V_N$ the nebular volume and
\be
\xi_B=\frac{3-p}{2-p} \frac{8 \pi\ m c^2}{V_N} \frac{a_2^{1/2}}{a_1} \frac{\nu_M^{\frac{2-p}{2}}-\nu_M^{\frac{2-p}{2}}}{\nu_M^{\frac{3-p}{2}}-\nu_M^{\frac{3-p}{2}}} S_{\rm TOT}
\label{eq:xib}
\ee
The equipartition field is then found simply equating the particles' and field energy:
\be
B_{\rm eq}=\xi_B^{2/7}
\label{eq:Beq}
\ee
while the minimum energy field is found by imposing $\partial W_{\rm TOT}/\partial B=0$, which gives:
\be
B_{\rm min}=\left(\frac{3}{4}\xi_B \right)^{2/7}\ .
\label{eq:Bmin}
\ee

For most astrophysical sources the assumption of equipartition is mostly made out of the need to obtain an estimate of the field strength, but in the case of PWNe there are also theoretical reasons to believe that equipartition is actually realised. We expect therefore that the strength of the magnetic field that we estimate based on equipartition or minimum energy (Eq.~\ref{eq:Beq} or \ref{eq:Bmin}) is not very different from that estimated from Eq.~\ref{eq:Bics}.

\section{Neutrino fluence estimate}
\subsection{Expected astrophysical events}
In this section we compute the neutrino fluxes that would be expected from the TeV detected PWNe in IceCube based on conversion of the whole photon flux in a corresponding number of neutrinos. We then discuss the most promising sources in detail and the implications of their non-detection by IceCube in terms of more refined predictions for ANTARES and KM3Net.
This will provide the most optimistic estimate for the hadronic contribution to the TeV photon fluence measured by the TeV telescopes.

Relativistic protons may produce TeV $\gamma$-rays either by photo-meson production or inelastic nuclear collisions. In \cite{Guetta} we showed that nuclear collisions are by far the most likely mechanism for pion production in PWNe.
The relation between the neutrino and photon flux is:
\begin{equation}
\int_{E_{\nu}^{\rm min}}^{E_{\nu}^{\rm max}}E_{\nu}\frac{dN_{\nu}}{dE_{\nu}}dE_{\nu}= \int_{E_{\gamma}^{\rm min}}^{E_{\gamma}^{\rm max}}E_{\gamma}\frac{dN_{\gamma}}{dE_{\gamma}}dE_{\gamma}
\end{equation}
where $ E_{\gamma}^{\rm min}$ ($E_{\nu}^{\rm min}$) and $E_{\gamma}^{\rm max}$ ($E_{\nu}^{\rm max}$) are the minimum and maximum photon (neutrino) energies respectively.

We estimate the neutrino flux in the energy range $1-100$ TeV, which is the range in which IceCube, and ANTARES  are operating and we assume the same range also for KM3Net.

The total number of expected astrophysical events in a year of operation of a neutrino telescope
is given by
\begin{equation}
N=\int_{1\,TeV}^{100\,TeV} 2\;T \frac{dN_{\gamma}}{dE_{\gamma}}A(E_{\nu},\delta)dE_{\nu} d\delta
\end{equation}
where $T$ is the exposure time of one year, $\frac{dN_{\gamma}}{dE_{\gamma}}$ is the TeV spectrum (described according to Eq.~\ref{eq:TeVspec} with the parameters given in Table \ref{tb:PWNlist}, according to the references reported in the table), $A(E_{\nu},\delta)$ is the effective area of the considered neutrino telescope, as a function of the neutrino energy $E_{\nu}$ and of the source declination, $\delta$. $A(E_{\nu},\delta)$ is shown in figures \ref{Iceeffarea}, \ref{ANTeffarea}, \ref{km3effarea}, for the IceCube, ANTARES and KM3NeT/ARCA detector respectively.
%...{\bf the ANTARES effective area is taken from the paper ApJ 760:53, 2012 November 20, the effective area is in Fig 6: m2 vs log10(E[GeV])\\  KM3Net ---letter of intent
%
%
%Irene
%bisogna mettere le aree efficaci in funzione dell'energia per ogni telescopio per favore aggiungi
%cosa hai fatto per stimare il numero di neutrini}.

\subsection{Expected atmospheric events}

The main component for the background is the flux of atmospheric neutrinos, which is caused by the interaction of cosmic rays, high energy protons and nuclei, with the Earth's atmosphere. Decay of charged pions and kaons produced in cosmic ray interactions generates the flux of atmospheric neutrinos and muons. Their energy spectrum is about one power steeper than the spectrum of the parent cosmic rays at Earth, due to the energy dependent competition between meson decay and interaction in the atmosphere. The spectral index for such power law is typically $\xi=2.7$. For the following estimates, we don't consider the additional atmospheric component due to the decay of heavier mesons, since this becomes relevant only for $E>100\;TeV$. 

The atmospheric neutrino flux is expressed as a power law
\begin{equation}
\frac{d\Phi_{\nu}}{dE_{\nu} d\Omega}=C_{\nu}E_{\nu}^{-\beta}
\end{equation}
where $C_{\nu}$ is a scale factor derived through Monte Carlo computations or experimental data, while $\beta \simeq \xi+1$.
The number of background neutrinos can be estimated as: 
\begin{equation}
BG=\int_{1\; TeV}^{100\; TeV} T \; \frac{d\Phi_{\nu}}{dE_{\nu} d\Omega} A(E_{\nu},\delta) dE_{\nu}\; d\delta\; d\Omega
\end{equation}
where $T$ is the exposure time of one year, $A(E_{\nu},\delta)$ is the relative effective area we already discussed, and $\frac{d\Phi_{\nu}}{dE_{\nu} d\Omega}$ is the atmospheric neutrino flux. The latter is estimated in \cite{bg} for ANTARES, and we use the same for KM3NeT/ARCA, whereas for IceCube we use the estimates given in \cite{Icebg}.

\section{Neutrinos from PWNe }
In this section we compute the number of neutrino events expected if the entire VHE $\gamma$-ray flux of TeV detected PWNe were of hadronic origin.

\begin{table} [h!]
\begin{center}
\begin{tabular}{|l|c|c|c|c|} \hline \hline 
Source &  $\delta$ & $N_0 \times 10^{11} $  & spectral  & reference\\
 Name & [$^\circ$] & $ [{\rm TeV}^{-1} {\rm cm}^{-2} {\rm s}^{-1} {\rm sr}^{-1 }$] & index & \\
\hline 
\hline 
Crab & 22.0145 & 2.8 & -2.6 & \cite{crab}\\
Vela & -45.17643 & 1.37 & -1.4 & \cite{vela}\\
G343.1-2.3 & -44.26667 & 0.23 & -2.5 & \cite{G343}\\
MSH15-52 & -59.1575 & 1.67 & -2.27 & \cite{MSH15}\\
G54.1+0.3 & 18.86667 & 0.075 & -2.39 & \cite{G54}\\
G0.9+0.1 & -28.15 & 0.084 & -2.4 & \cite{09}\\
G21.5-0.9 & -10.58333 & 0.046 & -2.08 & \cite{G21}\\
Kes75 & 2.983333 & 0.062 & -2.26 & \cite{G21}\\
J1356-645 & -64.5 & 0.27 & -2.2 & \cite{J1356}\\
CTA1 & 72.98361 & 0.102 & -2.2 & \cite{CTA1}\\
J1023-575 & -57.79 & 0.45 & -2.53 & \cite{J1023}\\
J1616-508 & -50.9 & 0.67 & -2.35 & \cite{J1616}\\
J1640-465 & -46.53 & 0.3 & -2.42 & \cite{J1616}\\
J1834-087 & -8.76 & 0.26 & -2.45 & \cite{J1616}\\
J1841-055 & -5.55 & 0.128 & -2.4 & \cite{J1841}\\
J1813-178 & -17.84 & 0.77 & -2.09 & \cite{J1616}\\
J1632-478 & -47.82 & 0.53 & -2.12 & \cite{J1616}\\
J1458-608 & -60.87722 & 0.21 & -2.8 & \cite{J1458}\\
J1420-607 & -60.76 & 0.35 & -2.17 & \cite{J1418}\\
J1809-193 & -19.3 & 0.46 & -2.2 & \cite{J1809}\\
J1418-609 & -60.97528 & 0.26 & -2.22 & \cite{J1418}\\
J1825-137 & -13.83889 & 0.198 & -2.38 & \cite{J1825}\\
J1831-098 & -9.9 & 0.11 & -2.1 & \cite{J1831}\\
J1303-631 & -63.1775 & 0.59 & -2.44 & \cite{J1303}\\
N 157B & -69.16583 & 0.13 & -2.8 & \cite{N157B}\\
J1837-069 & -6.95 & 0.5 & -2.27 & \cite{J1616}\\
J1912+101 & +10.15167 & 0.35 & -2.7 & \cite{J1912}\\
J1708-443 & -44.33333 & 0.42 & -2.0 & \cite{G343}\\
\hline 
\hline
\end{tabular}
\caption{\label{tb:PWNlist} The table contains the following fields as columns: the name of the source, its declination, $\delta$, in degrees, the number of 1 TeV photons per unit energy, unit surface and steradian detected from the source, $N_0$ (see Eq.~\ref{eq:TeVspec}), the $\gamma$-ray spectral index, $\alpha_{\gamma}$, and the reference from which the latter two values are taken.}
\end{center}
\end{table}
In Table~\ref{tb:ice} we report the total number of PWNe astrophysical events expected in one year of operation of the IceCube and ANTARES detectors, while in Table~\ref{tb:km3} the relative results are shown for KM3Net/ARCA. The number of atmospheric neutrino events collected in each detector is also reported. Since the dependence on declination of the effective area of KM3Net/ARCA is not yet available, the BG values are the same for each source, in this case, and only depend on the neutrino flavour. The neutrinos possibly produced in PWNe would only be muon and electron ones, however due to the effective neutrino oscillations an equal flux of all three flavours is expected in the detector.

\begin{table} [ht!!!!]
\begin{tabular}{|l|cc|cc|} \hline \hline 
     &IceCube & &ANTARES & \\
Name      & N  & BG & N & BG\\
\hline 
\hline
Crab& 13.09& 0.04& 0.07 & 0.02 \\
 \hline 
Vela& 0.76& 3.1e-07& 3.08 & 0.06 \\
 \hline 
G343.1-2.3& 1.8e-3& 3.1e-07& 0.02 & 0.04 \\
 \hline 
MSH15-52& 9.0e-3& 3.2e-07& 0.12 & 0.06 \\
 \hline 
G54.1+0.3& 0.54& 0.04& 3.3e-3 & 0.02\\
 \hline 
G0.9+0.1& 0.10& 5.6e-3& 7.8e-3 & 0.04 \\
 \hline 
G21.5-0.9& 0.13& 5.6e-3& 9.6e-3 & 0.04 \\
 \hline 
Kes75& 0.60& 0.04& 3.8e-3 & 0.02 \\
 \hline 
J1356-645& 8.4e-3& 3.7e-06& 0.05 & 0.06\\
 \hline 
CTA1& 0.89& 0.04& 0.02 & 0.06 \\
 \hline 
J1023-575& 3.2e-3& 3.2e-07& 0.04 & 0.06 \\
 \hline 
J1616-508& 9.3e-3& 3.2e-07& 0.09 & 0.06 \\
 \hline 
J1640-465& 3.2e-3& 3.2e-07& 0.04 & 0.06 \\
 \hline 
J1834-087& 0.28& 5.6e-3& 0.02 & 0.04 \\
 \hline 
J1841-055& 1.53& 5.64e-3& 0.12 & 0.04 \\
 \hline 
J1813-178& 20.63& 5.6e-3& 1.56 & 0.04 \\
 \hline 
J1632-478& 0.018& 3.2e-07& 0.13 & 0.06 \\
 \hline 
J1458-608& 6.9e-4& 3.7e-06& 0.01 & 0.06 \\
 \hline 
J1420-607& 0.01& 3.7e-06& 0.07 & 0.06 \\
 \hline 
J1809-193& 0.9& 5.6e-3& 0.07 & 0.04 \\
 \hline 
J1418-609& 7.5e-3& 3.7e-06& 0.05 & 0.06 \\
 \hline 
J1825-137& 2.5& 5.6e-3& 0.19 & 0.04 \\
 \hline 
J1831-098& 0.29& 5.6e-3& 0.02 & 0.04 \\
 \hline 
J1303-631& 7.4e-3& 3.7e-06& 0.07 & 0.06 \\
 \hline 
N 157B& 4.3e-4& 3.7e-06& 7.3e-3 & 0.06 \\
 \hline 
J1837-069& 0.836& 5.6e-3& 0.06 & 0.04 \\
 \hline 
J1912+101& 1.366& 0.04& 7.4e-3 & 0.02 \\
 \hline 
J1708-443& 0.02& 3.2e-07& 0.11 & 0.04 \\
 \hline  
\hline
\end{tabular}
\caption{\label{tb:ice} The table reports the predicted neutrino events from each of the considered PWNe in one year of operation of IceCube and ANTARES. The first column lists the names of the sources; the second one reports the number of expected astrophysical neutrinos (left) and background neutrinos (right) for IceCube; the third one is the same as the second, but for ANTARES. The average angular resolution detector is taken to be $0.6^{\circ}$ \cite{IcePoint} for the IceCube detector, and $0.46^{\circ}$ for ANTARES detector \cite{ANT}. The expected background events are computed in a sky patch of $3^{\circ} x \ 3^{\circ}$ for both detectors.}
\end{table}

\begin{table} [ht!!!]
\begin{tabular}{|l|ccc|} \hline \hline 
    & KM3Net/ARCA & &  \\
Name  & $N_{\nu_{\mu}}$  & $N_{\nu_{e}}$ & $N_{\nu_{\tau}}$\\
\hline 
\hline
Crab &25.29& 8.51 &9.46 \\
 \hline 
Vela &249.89& 58.43 &77.03 \\
 \hline 
G343.1-2.3 &2.48& 0.81 &0.91 \\
 \hline 
MSH15-52 &12.41& 4.05 &4.56 \\
 \hline 
G54.1+0.3 &0.99& 0.31 &0.36 \\
 \hline 
G0.9+0.1 &1.09& 0.35 &0.39 \\
 \hline 
G21.5-0.9 &1.21& 0.34 &0.41 \\
 \hline 
Kes75 &1.08& 0.33 &0.38 \\
 \hline 
J1356-645 &5.37& 1.59 &1.86 \\
 \hline 
CTA1 &2.03& 0.60 &0.70 \\
 \hline 
J1023-575 &4.60& 1.51 &1.70 \\
 \hline 
J1616-508 &9.67& 3.00 &3.45 \\
 \hline 
J1640-465 &3.77& 1.11 &1.36 \\
 \hline 
J1834-087 &3.08& 0.99 &1.12\\
 \hline 
J1841-055 &16.72& 5.28 &6.02\\
 \hline 
J1813-178 &197.84& 56.26 &67.12 \\
 \hline 
J1632-478 &12.68& 3.64 &4.33 \\
 \hline 
J1458-608 &1.38& 0.49 &0.53 \\
 \hline 
J1420-607 &7.46& 2.18 &2.57 \\
 \hline 
J1809-193 &9.157& 2.70 &3.17 \\
 \hline 
J1418-609 &4.95& 1.47 &1.72\\
 \hline 
J1825-137 &26.91& 8.44 &9.64 \\
 \hline 
J1831-098 &2.76& 0.79 &0.94 \\
 \hline 
J1303-631 &7.13& 2.28 &2.59 \\
 \hline 
N 157B &0.85& 0.30 &0.33 \\
 \hline 
J1837-069 &8.54& 2.58 &2.99 \\
 \hline 
J1912+101 &2.68& 0.93 &1.02 \\
 \hline 
J1708-443&13.46& 3.72 &4.50\\
 \hline  
\hline
\end{tabular}
\caption{\label{tb:km3}This table shows the results for the KM3Net/ARCA detector. The first column displays the name of the source, from the second to the fourth there are the number of expected astrophysical events as function of the different neutrino flavors $\nu_{\mu}$, $\nu_{e}$ and $\nu_{\tau}$, considering a nominal angular resolution of  $0.3^{\circ}$, \cite{IcePoint}. Since the effective area of KM3Net/ARCA is not given as function of the declination, the BG values are the same for each source, and are respectively: $BG_{\nu_{\mu}}=9.65$, $BG_{\nu_{e}}=4.63$, $BG_{\nu_{\tau}}=4.64$. The expected background events are computed in a sky patch of $3^{\circ} x \ 3^{\circ}$.}
\end{table}

The first thing that one notices looking at Table \ref{tb:ice} is that by now IceCube should have detected neutrinos from at least two sources, the Crab Nebula and J1813-178, if the entire $\gamma$-ray flux from those sources were of hadronic origin. Under the same assumption, a smaller, but still finite, number of events would have also been expected from a handful of other sources in 6 years of integration with IceCube. The latter sources notably include Vela, which, as one can see from Table \ref{tb:km3} and from the third column of Table \ref{tb:ice}, is the most promising candidate source for KM3Net and ANTARES. Indeed, while the number of neutrino events expected in IceCube is larger for J1813-178, when the computation is performed for the parameters appropriate for the next generation detectors (ANTARES and KM3Net), a factor of 2 more neutrinos are expected from Vela.

Further analyzing the tables \ref{tb:ice} and \ref{tb:km3}, one notices that additional promising neutrino sources  for ANTARES and KM3Net are, from top to bottom: MSH15-52, J1825-137 and J1841-055. Before further commenting these results, let us briefly discuss the properties of the six sources just mentioned, namely: Crab, Vela, MSH15-52, J1841-055, J1813-178, J1825-137, in the order in which they are listed in the table.

The first three sources had already been considered by \cite{Guetta} as promising candidates for neutrino detection. These are well studied PWNe for which multiwavelength data are available and spectral modelling has been developed by different authors. In Table~\ref{tb:Bfield} we report, for each of these three sources, the magnetic field strength as estimated based on two different arguments: in column 7 we report the value $B_{\rm eq}$ that can be estimated based on the equipartition argument (Eq.~\ref{eq:Beq}) applied to their X-ray emission properties (summarized in columns from 2 to 6 of the same table); in column 8 we report the field strength that can be estimated based on spectral modelling, $B_{\rm ICS}$ (either through Eq.~\ref{eq:Bics}, or based on more sophisticated modelling as we discuss below). 

\begin{table} [ht!!!]
\begin{center}
\begin{tabular}{|l|c|c|c|c|c|c|c|} \hline \hline 
Source &  $L_X$ & $\epsilon_1$ & $\epsilon_2$ & $\alpha_X$ & $R_{\rm PWN}$ & $B_{\rm eq}$ & $B_{\rm ICS}$\\
 Name & [$10^{35}$ erg/s] &  [keV]& [keV] & & [pc] & $\mu$G & $\mu$G\\ 
\hline 
\hline 
Crab & 200 & 0.5 & 8 & 1.12 & 1.2 & 200 & 150\\ 
Vela & $1.3 \times 10^{-3}$ & 0.5 & 8 & 0.4 & 0.1 & 25 & 5\\
MSH15-52 &0.4 & 0.5 & 8 & 0.65 & 4.5 & 20& 15 \\
\hline 
\hline
\end{tabular}
\caption{\label{tb:Bfield} The table contains the following fields as columns: the name of the source (first column); its X-ray luminosity (second column) integrated in the photon energy interval $\epsilon_1$-$\epsilon_2$ as specified in column 3 and 4; the spectral index (column 5) appropriate to describe the X-ray emission as $F_\nu \propto \nu^{-\alpha_X}$ with $F_\nu$ the energy emitted per unit frequency, time, surface and steradian; the X-ray radius of the nebula (column 6) approximated as a sphere; the equipartition magnetic field (column 7) estimated based on the specified X-ray emission properties; the strength of the magnetic field derived from spectral modelling, assuming that the $\gamma$-ray emission is all due to ICS (column 7). See the text for further details.}
\end{center}
\end{table}

In the following we discuss each of these sources in some detail.
\subsection{Crab}
\label{sec:crab}
The Crab Nebula is the typical PWN and one of the best studied objects in the sky. It is a very bright source of photons at all energies, and used to be considered a calibration source for all high energy telescopes, from X-rays to TeV $\gamma$-rays. The modelling of this source is obviously very well developed and it is the only PWN so far for which spatially resolved modelling of both the dynamics and emission properties is available at all frequencies (see \cite{amato14} for a review). A peculiarity of this source, is that it is such bright synchrotron emitter at all frequencies, that its synchrotron radiation is a non-negligible target photon field for ICS. The field estimate obtained from Eq.~\ref{eq:Bics} is however not far from what more sophisticated modelling provides (see e.g.~\cite{dejager}). This value is seen to be not far from the equipartition value and the $\gamma$-ray spectrum of the source is reasonably well accounted for within a pure leptonic scenario (see e.g.~\cite{bucciantini11, meyer10}). 

It is appropriate to point out, however, that having the Crab Nebula's $\gamma$-ray emission well explained as ICS does not mean that this source does not contain any relativistic hadrons, and not even that hadrons cannot be energetically dominant in the Crab pulsar wind. In fact, the latter condition could still be verified and the fact that we do not have any direct evidence for their presence in the $\gamma$-rays could be due either to the lack of target for p-p scattering, or just to the fact that the leptonic emission is overwhelming. In this latter case neutrino detection could still be possible and provide the only available test for the presence of hadrons in the source. 

The fact that no neutrinos have been detected from the Crab Nebula by IceCube implies that only $\approx 2 \%$ of the TeV $\gamma$-rays coming from this source can derive from p-p scattering, namely un upper limit on the flux of TeV photons corresponding to $F_{\rm up}=6\times10^{{-13}}{\rm TeV}^{-1}{\rm s}^{-1}{\rm cm}^{-2}{\rm sr}^{-1}$. This fact can be used, in principle, to put some constraints on the hadronic content of the Crab pulsar wind. This is what we will try to do in the following.

At the very high energies we are interested in, $E_{\gamma}>1$ TeV, the spectra of secondaries deriving from p-p scattering can be derived in the so-called scaling approximation (see e.g. \cite{berezinsky}):
\be
\frac{dN_{\gamma}}{dE_{\gamma} dt}=2c n_{H}\sigma_{pp}\int_{E_{\nu}}^{\infty}\frac{dE_{\pi}}{E_{\pi}}\int_{E_{\pi}}^{\infty}\frac{dE}{E_{\pi}}f\left(E_{\pi}/E\right)N_{p}(E)
\label{eq:dng0} 
\ee
where $f(x)=(1-x)^{3.5}+0.75\exp(-18x)$, $n_{H}$ is the density of target material, $\sigma_{pp}=5 \times 10^{-26}{\rm cm}^{2}$ is the total cross section for inelastic nuclear interactions. $E_{\pi}$ is the pion energy and $N_{p}(E)$ is the proton energy spectrum. 

In the framework of a cold pulsar wind and of termination shock that only accelerates pairs, the simplest assumption one can make about the proton energy spectrum is that of a mono energetic spectrum at $E_{w}=m_{i} \Gamma_{w}c^{2}$. On the other hand, depending on the physical processes that are at work at the termination shock to guarantee the acceleration of $e^{+}-e^{-}$ pairs, it cannot be excluded that wind ions might undergo further acceleration (or even deceleration). Therefore, in the following, we also consider a scenario in which the protons have a power-law spectrum. In both cases, the normalisation of the spectrum will be such that the total number of protons currently contained in the nebula is 
\be
\int_{E_{\rm min}}^{E_{\rm max}}N_{p}(E)dE =\dot N_{{\rm GJ}}t_{\rm c}
\label{eq:pnorm}
\ee
where we have made the assumption that particle injected in the PWN during time $t_{c}$ are still confined in the nebula and the approximation that $\dot N_{{\rm GJ}}\sim const$ during this time. 

In reality $\dot N_{\rm GJ}$ will vary with time due to the dependence on the pulsar period shown in Eq.~\ref{eq:ngj}. However, the Crab pulsar has a well-known spin down law $P(t)=P_{0}(1+t/\tau)^{0.57}$ \cite{groth75}, with $P_{0}$ the initial period and $\tau=730\ {\rm yr}$. During the entire lifetime of the nebula its period has only increased by a factor $\approx 1.5$ and $\dot N_{\rm GJ}$ has correspondingly decreased by a factor 2.5. 

As far as the estimate of the protons' confinement time, $t_{c}$, is concerned, the three time-scales that come about are: $t_{\rm age}$, the  age of the nebula; $t_{\rm pp}=(c n_{H} \sigma_{pp})^{-1}$, the time for energy losses, essentially due to p-p interactions; and $t_{\rm diff}(E)=3R_{N}^{2}/(c r_{L})$, the time it takes a particle with Larmor radius $r_{L}$ to diffuse out of the nebula (of radius $R_{N}$) according to Bohm's law. These three time-scales have a different relative importance during the history of the nebula, due to their different scaling with quantities such as the nebular radius and magnetic field. In order to obtain an estimate, we approximate the nebular expansion as occurring at a constant velocity: $R_{N}\propto t$. We assume the density of target material in the nebula to be $n_{H}=3 M_{\rm ej}/(4\pi R_{N}^{3})\propto t^{-3}$ where $M_{\rm ej}$ is the mass contained in the SN ejecta. And finally we approximate the nebular magnetic field to decrease in time as $B_{\rm neb}\propto t^{-1}$ (again a description appropriate for a time over which the pulsar input is about constant, see e.g. \cite{amato03}). For a current value of the average magnetic field strength $B_{\rm neb}(t_{\rm age})\approx 100 \mu G$, we obtain that at any time $t_{\rm esc}>t_{{\rm age}}$, and $t_{\rm pp}>t_{{\rm age}}$ for $t_{\rm age}\gtrsim t_{x}=200$ yr. As a result, being the current age of the Nebula $t_{\rm Crab}=10^{3}$ yr, the time $t_{c}$ that enters the normalisation in Eq.~\ref{eq:pnorm} will be $t_{c}=800$ yr.

We can then compute the flux of photons resulting from Eq.~\ref{eq:dng0} under the two above mentioned assumptions on the proton spectrum.
In the case when
\be
N_{p}(E)=\dot N_{\rm GJ}t_{c}\delta \left(E-m_{i}\Gamma_{w}c^{2}\right)\ ,
\label{eq:pdirac}
\ee
the minimum proton energy is $E_{\rm min}>10$ TeV, even for the smallest reasonable value of the wind Lorentz factor it is easy to show that one obtains:
\be
F^{\delta}_{\gamma}(E_{\gamma})=\frac{1}{4\pi d^{2}}\frac{dN_{\gamma}}{dE_{\gamma}dt}=\frac{2c n_{H}\sigma_{pp} \dot N_{\rm GJ}t_{c}}{4 \pi d^{2}E_{w}}\int_{\frac{E_{\gamma}}{E_{w}}}^{1}\frac{f(x)}{x^{2}}dx\ .
\label{eq:dngdirac}
\ee
The wind Lorentz factor $\Gamma_{w}$ is uncertain, and the fraction of pulsar wind energy that ions can carry, $f_{p}$, depends on its value as:
\be
f_{p}=\frac{\dot N_{\rm GJ}m_{i} c^{2}\Gamma_{w}}{L_{p}}\approx \frac{10^{7}}{A\Gamma_{w}}\ ,
\label{eq:fp}
\ee
with $A$ the mass number.
The lowest estimate of $\Gamma_{w}$ in the Crab Nebula corresponds to $\Gamma_{w}=10^{4}$ (see \cite{amato14} for a review), which translates in a negligible fraction of the pulsar wind energy carried by ions: $f_{p}<10^{-3}$. The largest possible Lorentz factor corresponds to the total potential drop available in the pulsar magnetosphere and ions carrying almost all of the wind energy: $\Gamma_{w}\approx10^{7}$. For large Lorentz factors, the the lower limit of integration in Eq.~\ref{eq:dngdirac} becomes small enough that one can approximate the integral as $\approx 1.5 E_{w}/E_{\gamma}$ and obtain the expected $1/E_{\gamma}$ scaling expected for the production of photons of energy much lower than the parent proton energy. The expected number of photons becomes independent of the parent proton energy: 
\be
F^{\delta}_{\gamma}(E_{\gamma},E_{w}\gg E_{\gamma})\approx\frac{3.5 c \sigma_{pp} \dot N_{\rm GJ}t_{c}}{E_{\gamma}}\ .
\label{eq:fdeltalarge}
\ee
For the values of the parameters that are appropriate for Crab one finds
\be
F^{\delta}_{\gamma}(1{\rm TeV},\Gamma_{w}>10^{4})\approx 
\label{eq:fdelta}
\ee
$$\approx 6 \times 10^{-14}\left(\frac{E_{\gamma}}{\rm 1TeV}\right)^{-1}\left(\frac{M_{\rm ej}}{10 M_{\odot}}\right){\rm TeV}^{-1}{\rm s}^{-1}{\rm cm}^{-2}{\rm sr}^{-1}
$$
The approximation valid for $E_{w}\gg E_{\gamma}$ cannot be used to compute the flux of TeV photons in the case $\Gamma_{w}=10^{4}$. In this case, the integral must be computed explicitly and the result is a factor of 6 lower flux. Of course the value of $F^{\delta}_{\gamma}$ depends linearly on the target density, which here we assumed to correspond to $10\ M_{\odot}$ of material uniformly distributed in the nebula. This assumption may be wrong in an unpredictable manner: as discussed in \cite{amato03} the material ejected in the SN event is concentrated in filaments penetrating the body of the nebula, rather than uniformly distributed; in addition we have no clue about the magnetic field structure in the proximity of these filaments and do not know whether trapping of energetic particles might occur, leading to increase the probability of interaction between the relativistic particles and the gas. The computed photon flux for $\Gamma_{w}>10^{4}$ is still about a factor 10 lower than the IceCube neutrinos can probe, but certainly within reach for the next generation neutrino telescopes with sensitivities increased by a factor 5-10. It is important to notice that these upcoming telescopes are likely to allow us to probe the presence of hadrons in PWNe independently of the wind Lorentz factor and hence, according to Eq.~\ref{eq:fp}, independently of the fraction of wind energy they carry, as long as this is larger than $10^{-3}$. 

The other major assumption underlying the calculation above is the fact that protons stay cold even after crossing the wind termination shock. For the sake of generality we compute the expected photon flux also for the case of a power-law distribution of relativistic hadrons: $N_{p}\propto E^{-\alpha_{p}}$ with $\alpha_{p}>2$ and $E_{\rm min}\lesssim E_{\gamma}$. In this case one has:
\be
F^{\rm pl}_{\gamma}(E_{\gamma})=\frac{2}{\kappa \alpha_{p}} \frac {c n_{H} \sigma_{pp}\dot N_{\rm GJ} t_{c}}{4 \pi d^{2}E_{\rm p,min}}\left(\frac{2E_{\gamma}}{\kappa}\right)^{-\alpha_{p}}\ .
\label{eq:dngth}
\ee
where $\kappa\approx 0.2$ is the average fraction of proton energy transferred to the pion and $E_{\rm p,min}$ is related to $E_{w}$. In the case of a power-law index $\alpha_{p}>2$ this relation is straightforward, with $E_{\rm p,min}\approx (\alpha_{p}-2)/(\alpha_{p}-1)E_{\rm w}$. For $\alpha_{p}<2$ an analogous relation holds between the wind energy and the maximum particle energy (rather than the minimum).
It is clear then that if the wind Lorentz factor is large ($\Gamma_{w}\gg10^{4}$) and $\alpha_{p}>2$, $E_{\rm p,min}\gg$1TeV and the flux of 1 TeV secondaries will be described again by Eq.~\ref{eq:fdelta}, even if the parent hadrons have a power-law distribution. On the other hand, for a wind Lorentz factor $\Gamma_{w}\approx 10^{4}$, if protons are power-law distributed, the predicted flux from the Crab Nebula is one order of magnitude lower than the expectation in Eq.~\ref{eq:fdelta}. 

More generally we find that the only case in which the flux of secondaries deriving from a power-law distribution of hadrons becomes comparable with that expected from a mono energetic distribution and close to the current detection limit is if most of the pulsar wind energy is carried by protons: $\gamma_{w}\approx 10^{6}$ and $f_{p}>20\%$.

\subsection{Vela}
\label{sec:vela}
The Vela PWN is the second best studied PWN. Data are again available at basically all frequencies, but their interpretation is far more complicated than for Crab because this is an old and very complex system, where the emission of the PWN proper is not easy to disentangle from other contributions (REF). Again a lot of modelling is available, but the nature of the TeV emission is in this case more controversial. For example \cite{hornsetal} suggested a most likely hadronic origin of the TeV emission from Vela, mostly based on energetic arguments, whereas \cite{dejagervela,lamassa08} proposed a fully leptonic model. In the light of the most recent release of Fermi data, \cite{grondinvela} showed that leptonic models can well account for the multi wavelength spectrum of the source with a magnetic field value of order 5 $\mu$G. This field strength is what we report as $B_{\rm ICS}$ in Table~\ref{tb:Bfield}. The estimate based on Eq.~\ref{eq:Bics} would give a much lower value, but this is not reliable for two reasons. First of all, if the magnetic field is so low as $\sim \mu$G the electrons responsible for ICS emission of TeV photons are not the ones radiating X-rays through synchrotron emission. In addition, in the case of Vela, according to the modelling of \cite{grondinvela}, the most important contribution to the target photon field for ICS comes from an enhanced IR background (a factor of 5 larger than the galactic average) that Eq.~\ref{eq:Bics} does not take into account. It is interesting to notice that an enhanced IR background is often invoked in the leptonic modelling of the $\gamma$-ray emission of PWNe, and actually SNRs in general (see e.g.~\cite{morlinorxj}): on one hand, a larger than average IR field is expected in sources that are thought to host large amounts of dust, on the other hand it is very unfortunate that it is exactly these sources that could show signatures of relativistic protons in the energy range where enhanced ICS is bound to cover them.

\subsection{MSH15-52}
\label{sec:msh}
MSH15-52 is the spectacular nebula produced by the pulsar PSR B1509-58, which is one of the youngest, most energetic pulsars known, with its 1700 yr of age, its 150 ms period and it estimated surface magnetic field of order $1.5 \times 10^{13}$ G \cite{kaspib1509}. The nebula has been observed in the radio and X-ray band \citep{gaenslermsh} and in GeV \citep{fermimsh} and TeV \citep{aharmsh} $\gamma$-rays. In spite of the morphological complexity of the system, thanks to the abundance of data, spectral modelling has been possible and the result is that the $\gamma$-ray flux of the source is well explained with a magnetic field value very close to the equipartition one (17 $\mu$G vs 22 $\mu$G, \cite{fermimsh}). The main contribution to the target photon field for ICS comes from the IR background, that in this case is taken to be at the average level in the Galaxy as inferred from GALPROP. It is important to mention that \cite{fermimsh} also attempted to interpret the $\gamma$-ray spectrum of this source as the result of $\pi^0$ decay: in spite of the fact that the source is in a rather dense region, the energy that is required to be put in relativistic protons in order to explain the entire $\gamma$-ray flux as hadronic, exceeds the total energy supplied by the pulsar. Once again this does not preclude the possibility that at least part of the high energy photons ore of hadronic origin, and in terms of abundance of target for nuclear collisions this is one of the most promising sources.

\subsection{J1841-055}
\label{sec:J1841}
This TeV source, initially discovered by \cite{J1841} in the HESS Galactic Plane Survey, is likely associated with PSR J1838-0537 \cite{pletsch}. A fraction of 0.05\% of the pulsar energy, similar to other pulsar-PWN systems, would be sufficient to explain the extended TeV emission. The Fermi $\gamma$-ray telescope found GeV emission from that region with a spectrum that nicely connects with the TeV data and results in a high energy SED similar to what is typical for PWNe \cite{acero13}. However the source is very extended, both in GeV and TeV $\gamma$-rays, so that it is not clear whether the entire photon flux comes from a single object. No detection at other wavelength is available. More data are needed for any further discussion.
 
\subsection{J1813-178}
\label{sec:J1813}
The TeV source J1813-178 was discovered in the HESS survey of the inner Galaxy \cite{ahar05Sci}. The source was initially identified as a SNR \citep{brogan}, but later \cite{funk1813}, based on XMM-Newton data, proposed that the complex morphology was better explained as a SNR-PWN system. Current spectral modelling \citep{funk1813}, taking into account all available data and Fermi upper limits, explains the TeV emission as due to ICS, with the target photon field contributed by CMB, IR and FIR background. The resulting magnetic field is $B\approx 7 \mu$G, which is of order of the equipartition magnetic field that can be estimated based on the X-ray data associated with the nebula \citep{funk1813}. Once again a purely leptonic origin of the $\gamma$-ray emission seems plausible.

\subsection{J1825-137}
\label{sec:J1825}
This TeV source, initially found by HESS \citep{J1825} and later also seen by Fermi \citep{grondinJ1825} is associated with an old PWN displaced from its parent pulsar position. As discussed by \cite{grondinJ1825} (see also references therein), the high energy emission is more extended than the compact X-ray core of the source and the high energy emission seems well interpreted within a leptonic model with a magnetic field $B\approx 3-4 \mu$G.  The estimated energy in the form of electrons is $W_e\approx 5 \times 10^{49}$ erg and for a size of order $1^\circ$ a comparable energy would be associated to the inferred magnetic field. Once again the leptonic scenario gives reasonable results.

 \section{Revised neutrino predictions in the light of IceCube results}
In this section we revise our prediction of the neutrino flux to be expected from our most promising sources in the light of their non-detection by IceCube. For some of the sources discussed above, the fact that IceCube has detected no neutrinos in 6 years of integration implies, independently of any theoretical consideration, that the detected TeV $\gamma$-ray flux cannot be entirely of hadronic origin. Since, as can be seen from Table~\ref{tb:ice}, the expected neutrino counts in one yr of operation of IceCube are in all cases well above the background, we simply revise our estimate of the neutrino counts expected in ANTARES and KM3Net/ARCA by dividing the counts listed in column 3 of Table~\ref{tb:ice} (left hand side) and in the three columns of Table~\ref{tb:km3} by $\max[1,6N_{{\rm ICE}}]$ with $N_{\rm ICE}$ the counts listed in the second column of Table~\ref{tb:ice} (left hand side). The results are reported in Table~\ref{tb:nurev}.

\begin{table} [ht!!!]
\begin{center}
\begin{tabular}{|c||c||ccc||} \hline \hline 
 &  ANTARES & &KM3Net/ARCA & \\
 Source & $N_{\nu}$ & $N_{\nu_{\mu}}$ & $N_{\nu_{e}}$ & $N_{\nu_{\tau}}$\\
\hline 
\hline 
Crab & $10^{-4}$ & 0.3 & 0.1 & 0.12\\ 
Vela & 0.7 & 54.8 & 12.8 & 16.9\\
MSH15-52 & 0.12 & 12.4 & 4 & 4.6\\
J1841-055 & 0.01 & 1.8 & 0.58 & 0.66\\
J1813-178 & 0.01 & 1.6 & 0.45 & 0.54\\
J1825-137 & 0.01 & 1.8 & 0.56 & 0.64\\
\hline 
\hline
\end{tabular}
\caption{\label{tb:nurev} The table contains the following fields as columns: the name of the source (first column); the revised expectation of neutrino counts for 1 yr of integration of ANTARES (column 2); the revised expectation for the different kinds of neutrinos for KM3Net/ARCA.}
\end{center}
\end{table}
It is clear from the Table that the most promising PWN to be detected by upcoming neutrino telescopes is Vela. At the same time, detection of neutrinos from MSH15-52 also appears possible.

\section{Summary and Conclusions}
We have investigated the implications of a possible hadronic origin of the high energy emission from the PWNe which have been detected by H.E.S.S. at TeV energies \cite{site}. The alternative explanation for this emission is purely leptonic (based on ICS). In reality both leptons and hadrons could contribute to the emission, and the relative contribution affects the predictions for upcoming neutrino telescopes. Interesting constraints come, for some of the sources, from the non-detection by IceCube.

We devoted special attention to six sources from which the neutrino number counts in upcoming detectors were found to be especially high, when assuming that their entire TeV photon flux had a hadronic origin. These sources are: Crab, Vela, MSH15-52, J1841-055, J1813-178, J1825-137. 
In the case of the Crab Nebula, it is clear that the TeV $\gamma$-ray flux is mostly contributed by ICS. The existing upper limits on the neutrino flux derived from IceCube non-detection do not allow us to put constraints on the hadronic content of the Crab pulsar wind. However, with the next generation neutrino telescopes we will be likely able to probe the presence of relativistic hadrons in the Crab Nebula: with a factor of ten larger sensitivity, TeV neutrinos should be detectable from Crab, in 5-10 years of integration with KM3Net/ARCA, almost independently of the pulsar wind parameters, in the case of cold hadrons; they should also be detectable for a power law distribution of hadrons, if these carry more than 20\% of the pulsar wind energy.

The second source we considered, Vela, turns out to be the best candidate PWN to be detected in neutrinos. Even taking into account the IceCube non-detection, the revised number of expected neutrinos still suggests that the source can be detected by ANTARES in a few years of integration and promptly by KM3Net/ARCA. We also mentioned the difficulties associated with a fully leptonic interpretation of the $\gamma$-ray flux from this source: namely the requirement that the nebular magnetic field be on average a factor 5 below the estimated equipartition field and the presence of an enhanced IR background. Taking into account all this, Vela really appears to be a promising neutrino source.

A detectable neutrino flux is also expected from MSH15-52, while for the remaining potentially promising sources in our list, the IceCube constraints strongly reduce the perspectives of detection with the KM3Net detector.

It is important to note that the KM3Net/ARCA predictions were derived without considering the effect of the efficiency of the apparatus and the selection and reconstruction criteria.
% While ANTARES is better suited than IceCube to detect neutrino events from PWNe that have negative declinations, the IceCube non detection does not leave much hope for detecting PWNe with this telescopes. ANTARES is not expected to see any PWN in less than 10 years of integration, with the possible exception of Vela. Indeed, if the entire $\gamma$-ray flux from this source were of hadronic origin, not only Vela should have been detected by IceCube, but 3 neutrinos should have also been observed in ANTARES. Our revised estimate explains the lack of an excess in that direction.
%
%The perspectives of detection remain much better for KM3Net/ARCA. 

The Cherenkov Telescope Array will likely increase the number of PWNe detected at TeV energy to several hundreds, probably providing an essentially complete account of TeV emitting PWNe in the Galaxy. The analysis performed in this work can be easily extended to each upcoming TeV detected PWN.

\end{document}